\documentclass[prd,amsmath,amssymb,showpacs,superscriptaddress,nofootinbib]{revtex4}
\usepackage{amssymb}
\usepackage{graphicx,epsfig}
\usepackage{dcolumn}
\usepackage{bm}
\usepackage{psfig}
\usepackage[dvipdfm,CJKbookmarks=true,unicode,colorlinks,linkcolor=blue,anchorcolor=blue,citecolor=blue,pdfborder={0 0 0}]{hyperref}
\usepackage{amsfonts}
\usepackage{keyval,graphicx}
\usepackage{textcomp,wasysym}

\begin{document}

\normalsize
\parskip=5pt plus 1pt minus 1pt

\title{\boldmath Observation of $\chi_{cJ}$ decaying into the $p\bar{p}K^{+}K^{-}$ final state}

\author{
{\small M.~Ablikim$^{1}$, M.~N.~Achasov$^{5}$, D.~Alberto$^{38}$, L.~An$^{9}$, Q.~An$^{36}$, Z.~H.~An$^{1}$, J.~Z.~Bai$^{1}$, R.~Baldini$^{17}$, Y.~Ban$^{23}$, J.~Becker$^{2}$
% ...
\\
\vspace{0.2cm}
(BESIII Collaboration)\\
\vspace{0.2cm} {\it
$^{1}$ Institute of High Energy Physics, Beijing 100049, P. R. China\\
$^{2}$ Bochum Ruhr-University, 44780 Bochum, Germany\\
% ...
$^{5}$ G.I. Budker Institute of Nuclear Physics SB RAS (BINP), Novosibirsk 630090, Russia\\
% ...
$^{9}$ Guangxi University, Naning 530004, P. R. China\\
% ...
$^{17}$ Laboratori Nazionali di Frascati - INFN, 00044 Frascati, Italy\\
% ...
$^{23}$ Peking University, Beijing 100871, P. R. China\\
% ...
$^{36}$ University of Science and Technology of China, Hefei 230026, P. R. China\\
% ...
$^{38}$ University of Turin and INFN, Turin, Italy\\
% ...
}}
\vspace{0.4cm}
}

\begin{abstract}

First measurements of the decays of the three $\chi_{cJ}$ states to
$p\bar{p}K^{+}K^{-}$ final states are presented.  Intermediate $\phi\to
K^{+}K^{-}$ and $\Lambda(1520)\to pK^{-}$ resonance states are observed, and
branching fractions for $\chi_{cJ}\to  \bar{p}K^{+}\Lambda(1520)$,
$\Lambda(1520) \bar{\Lambda}(1520)$, and $\phi p\bar{p}$ are reported.
We also measure branching fractions for direct $\chi_{cJ}\to p\bar{p}
K^{+}K^{-}$ decays.  These are first observations of $\chi_{cJ}$ decays to
unstable baryon resonances and provide useful information about the
$\chi_{cJ}$ states. The experiment uses samples of $\chi_{cJ}$ mesons
produced via radiative transitions from 106 million $\psi^{\prime}$ mesons
collected in the BESIII detector at the BEPCII $e^+e^-$  collider.

\end{abstract}

\pacs{13.25.Gv, 14.20.Pt, 14.40.Be}

\maketitle

\section{Introduction}

Experimental studies on charmonia decay properties are useful for
testing perturbative QCD models and QCD-based calculations. In the
standard quark model, the $\chi_{cJ}(J=0,1,2)$ mesons are P-wave
quarkonium states with spin parity $0^{++}$, $1^{++}$ and
$2^{++}$. Although they cannot be produced directly in $e^{+}e^{-}$
collisions, radiative decays of the $\psi^\prime$ into each
$\chi_{cJ}$ occur about $9\%~$\cite{PDG} of the time and provide
large $\chi_{cJ}$ samples that have proven to be a very clean
environment for studies of the $\chi_{cJ}$ states.

The color octet mechanism (COM) has been shown to play an important
role in describing these P-wave quarkonium
decays~\cite{SMW_NUL,JPG_PLB,SMW_NPA,EJ_PRD}.  Many COM predictions
for $\chi_{cJ}$ decays into meson pairs and $p \bar{p}$ pairs are in
agreement with earlier experimental results.  However, the predictions
for some baryon-antibaryon decays disagree with measured values, in
particular
$\chi_{cJ}\rightarrow\Lambda\bar{\Lambda}$~\cite{JZB_RPD}. At present,
only ground state baryons have been observed in $\chi_{cJ}$
decays~\cite{PDG}.  To further test COM predictions for P-wave
charmonia decay, measurements of excited baryon pair decays are
important. This paper presents a study of $\chi_{cJ}$ hadronic decays
and measurements of
$\chi_{cJ}\rightarrow\Lambda(1520)\bar{\Lambda}(1520)$ decaying to
$p\bar{p}K^{+}K^{-}$, based on $106$ million $\psi^\prime$ events
collected with BESIII at BEPCII.  The observation of such excited
baryon production can provide constraints on models of P-wave
charmonia hadronic decay.

\section{Detector}

BEPCII~\cite{NIM_DET} is a double-ring $e^{+}e^{-}$ collider designed
to provide a peak luminosity of $10^{33}$$~$ cm$^{-2}s^{-1}$ at the
center of mass energy of $3770$ MeV. The BESIII~\cite{NIM_DET}
detector has a geometrical acceptance of $93\%$ of $4\pi$ and has four
main components: (1) A small-cell, helium-based ($40\%$ He, $60\%$
C$_{3}$H$_{8}$) main drift chamber (MDC) with $43$ layers providing an
average single-hit resolution of $135$ $\mu$m, and charged-particle
momentum resolution in a $1$ T magnetic field of $0.5\%$ at $1$
GeV$/c$. (2) An electromagnetic calorimeter (EMC) consisting of $6240$
CsI(Tl) crystals in a cylindrical structure (barrel) and two
endcaps. The energy resolution at $1.0$ GeV$/c$ is $2.5\%$ ($5\%$) in
the barrel (endcaps), and the position resolution is $6$ mm ($9$ mm)
in the barrel (endcaps). (3) Particle Identification is provided by a
time-of-flight system (TOF) constructed of $5$-cm-thick plastic
scintillators, with $176$ detectors of $2.4$ m length in two layers in
the barrel and $96$ fan-shaped detectors in the endcaps. The barrel
(endcap) time resolution of $80$ ps ($110$ ps) provides $2\sigma$
$K/\pi$ separation for momenta up to $\sim 1.0$ GeV$/c$.  (4) The muon
system (MUC) consists of $1000$ m$^{2}$ of Resistive Plate Chambers
(RPCs) in nine barrel and eight endcap layers and provides $2$ cm
position resolution.

\section{Monte-Carlo simulation}
Monte-Carlo (MC) simulation of the full detector is used to determine
the detection efficiency of each channel, optimize event selection
criteria, and estimate backgrounds. The simulation program, BOOST,
provides an event generator, contains the detector geometry
description, and simulates the detector response and signal
digitization.  Charmonium resonances, such as the $\psi^\prime$, are
generated by KKMC~\cite{KKMC_1,KKMC_2}, which accounts for effects
such as initial state radiation and beam energy spread. The subsequent
charmonium meson decays are produced with
BesEvtGen~\cite{Evtgen_1,Evtgen_2}. The detector geometry and material
description and the tracking of the decay particles through the
detector including interactions are handled by Geant4.

\section{Event Selection}

Charged tracks must have their point of closest approach to the
beamline within $\pm 10$ cm of the interaction point in the beam
direction and within $1$ cm of the beamline in the plane perpendicular
to the beam and must have the polar angle satisfy $|\cos\theta|<0.93$.
The TOF and energy loss dE/dx measurements are combined to calculate
particle identification (PID) probabilities for pion, kaon, and
proton/anti-proton hypotheses, and each track is assigned a particle
type corresponding to the hypothesis with the highest confidence level
(C.L.). Finally, four tracks identified as $p$, $\bar{p}$ $K^{+}$, and
$K^{-}$ are required.

Photon candidates are selected by requiring a minimum energy
deposition of $80$ MeV in the EMC. EMC cluster timing requirements
suppress electronic noise and energy deposits unrelated to the event.

Kinematic fitting that utilizes momentum and energy conservation is
applied under the hypothesis
$\psi^\prime\rightarrow\gamma\chi_{cJ}\rightarrow\gamma
p\bar{p}K^{+}K^{-}$.  For events with more than one photon candidate,
the combination with the smallest $\chi^{2}_{4C}$ is considered for
further analysis.

\section{Data Analysis}
\begin{figure*}[htbp]
   \centerline{
   \psfig{file=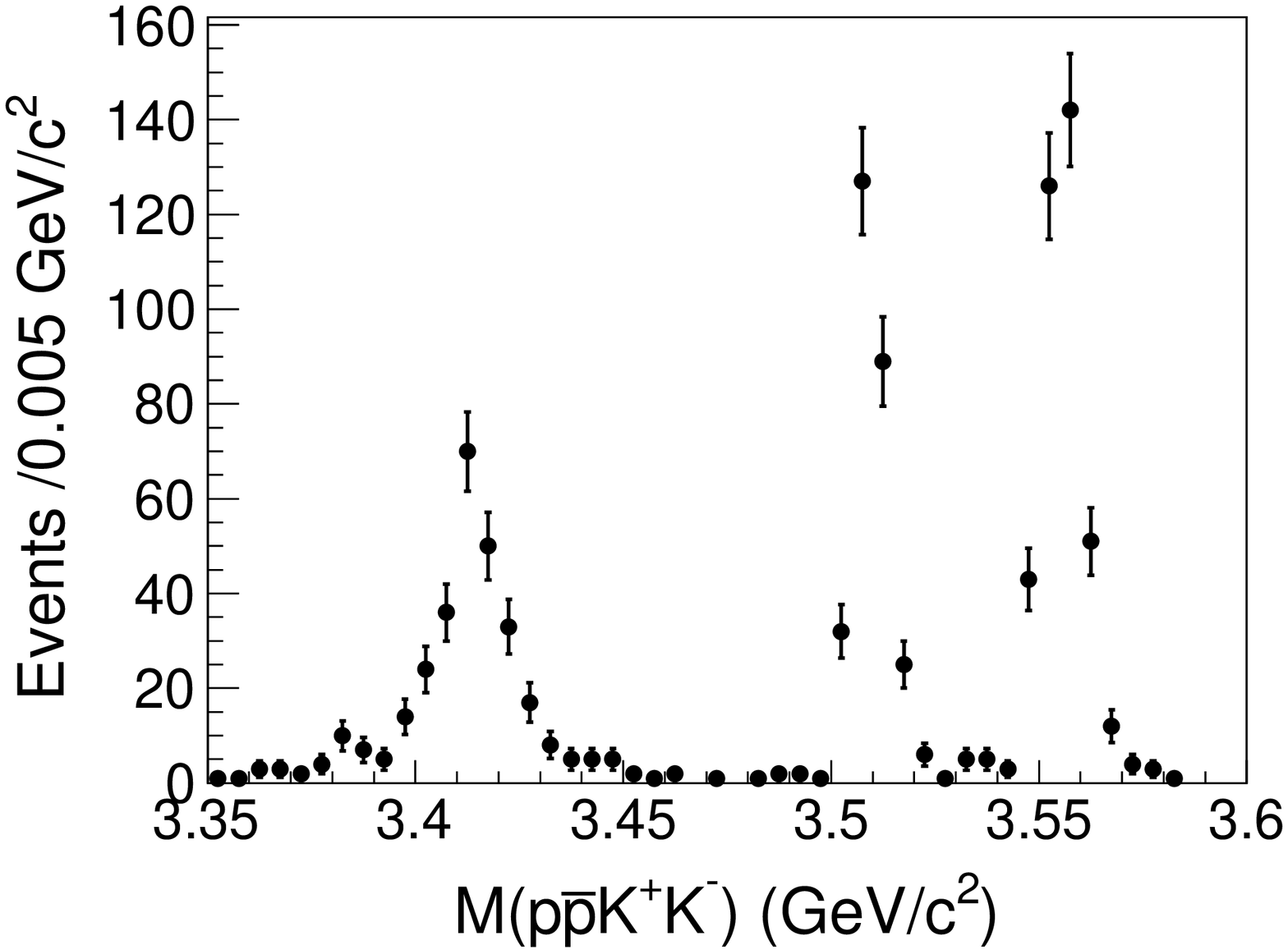,width=7cm,height=6cm, angle=0}
              \put(-35,135){(a)}
   \psfig{file=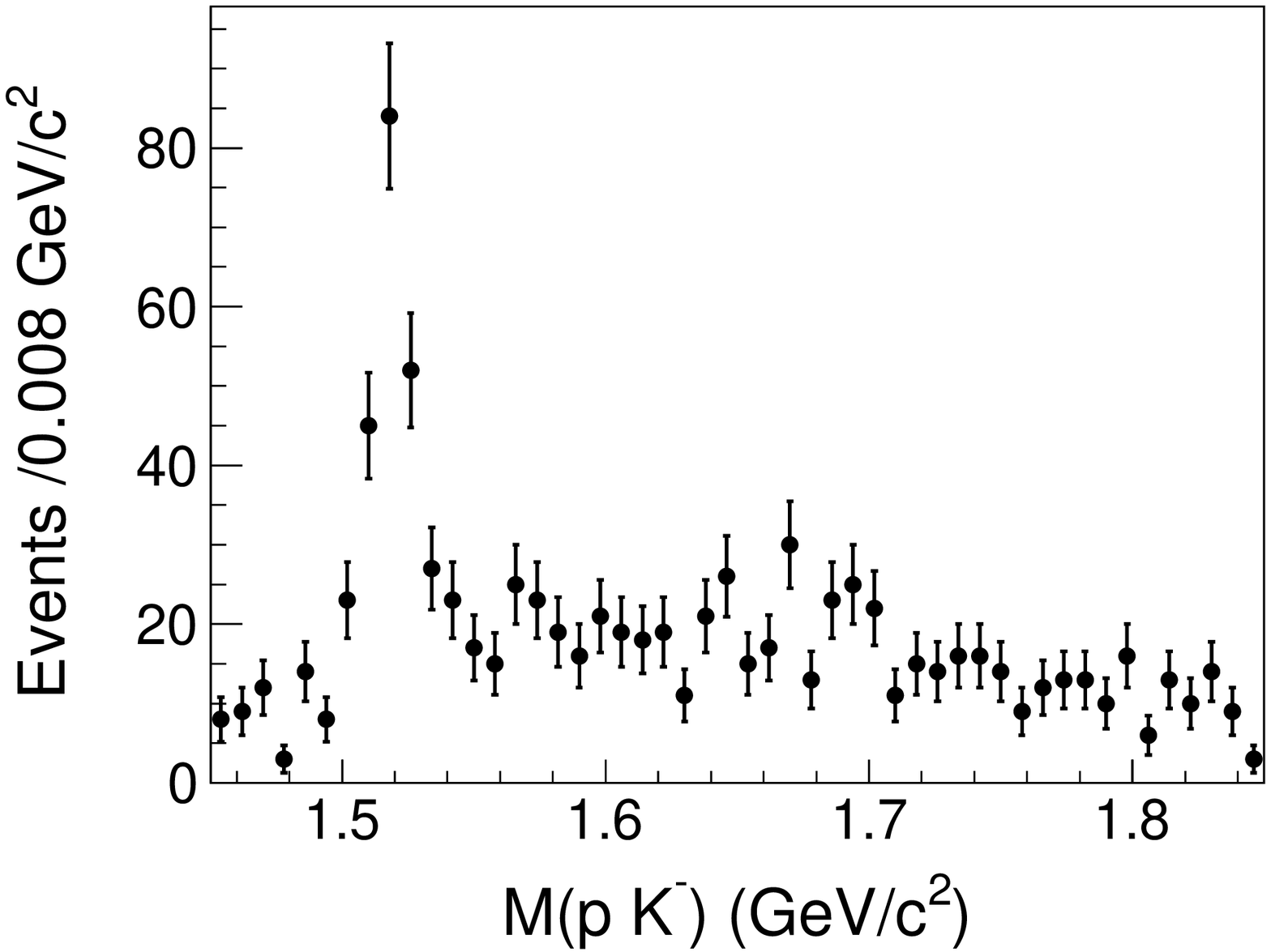,width=7cm, height=6cm, angle=0}
              \put(-35,135){(b)}}
   \centerline{
   \psfig{file=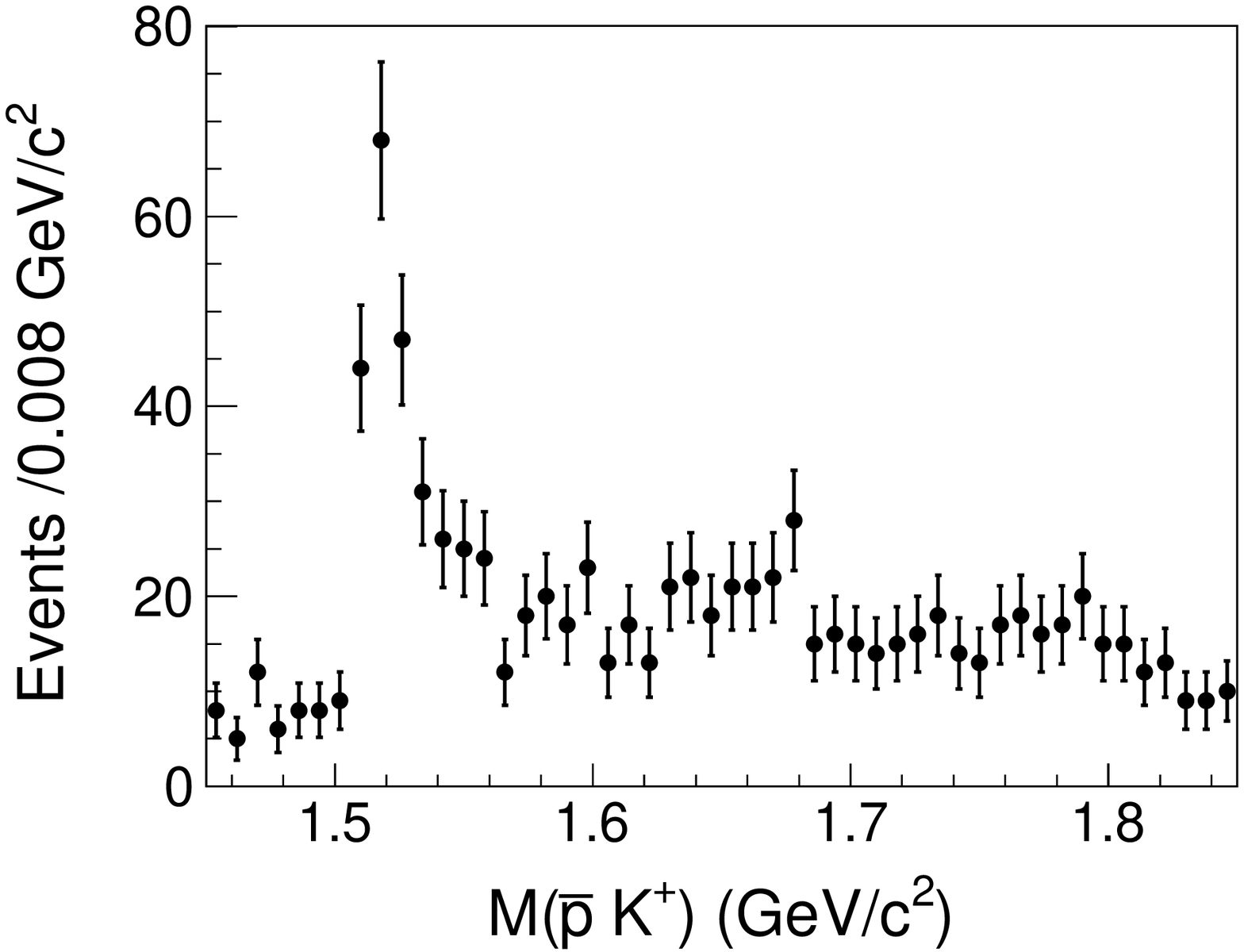,width=7cm,height=6cm, angle=0}
              \put(-35,135){(c)}
   \psfig{file=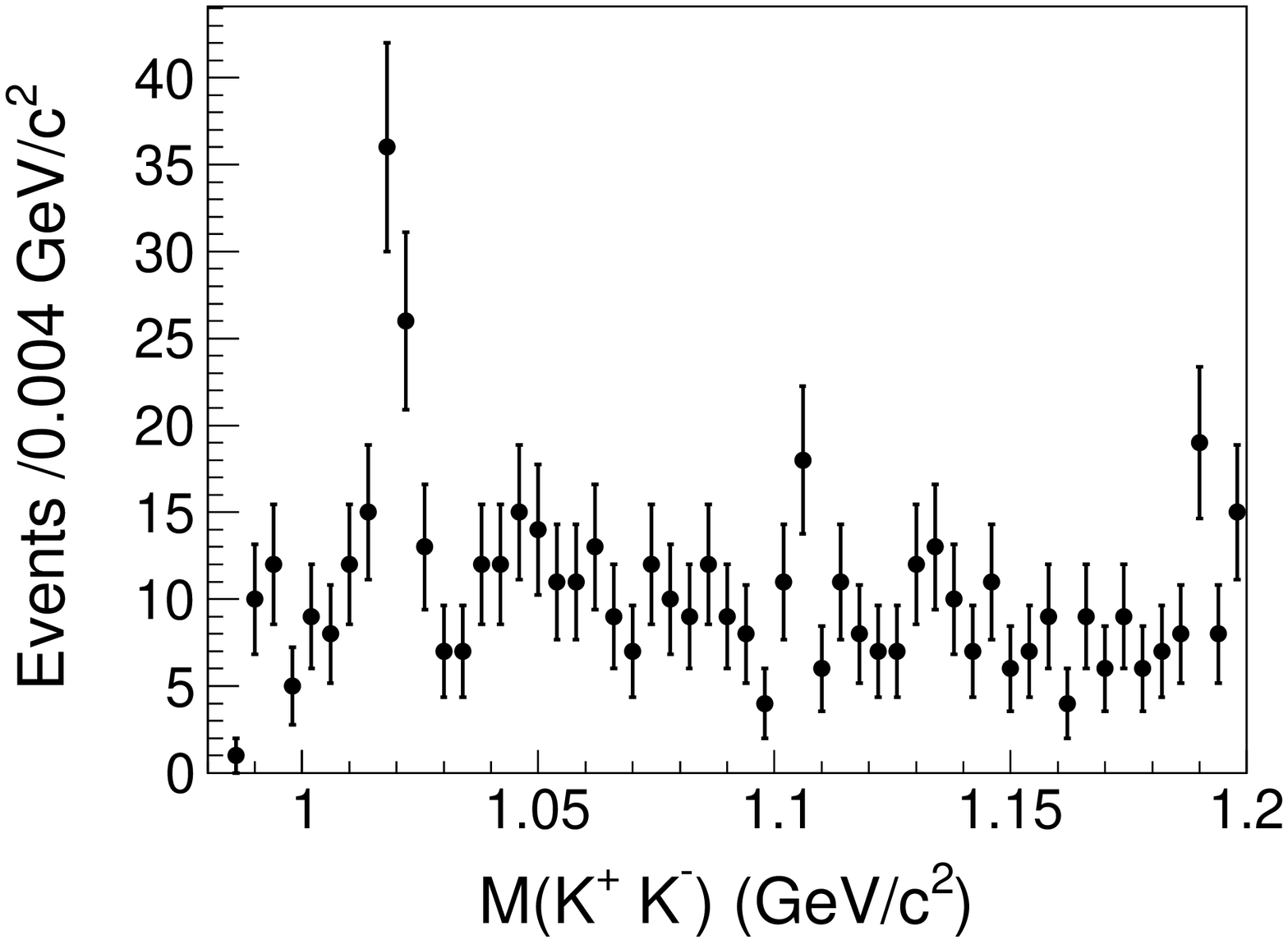,width=7cm, height=6cm, angle=0}
              \put(-35,135){(d)}}
   \caption{Invariant mass distributions of (a) $p\bar{p}K^{+}K^{-}$, (b) $pK^{-}$,
            (c) $\bar{p}K^{+}$ and (d) $K^{+}K^{-}$}.
   \label{fig:mass}
\end{figure*}

After candidate event selection, distinct $\chi_{cJ}$ signals are
observed in the $p\bar{p}K^{+}K^{-}$ invariant mass distribution, as
shown in Fig.~\ref{fig:mass}(a). By combining final state particles
($p,\bar{p},K^{+},K^{-}$), the $\Lambda(1520)$, $\bar{\Lambda}(1520)$
and $\phi$ intermediate states can also be seen in the $pK^{-}$,
$\bar{p}K^{+}$ and $K^{+}K^{-}$ invariant mass distributions, as shown
in Fig.~\ref{fig:mass}(b,c,d), respectively.

\subsection{Background Studies}

A sample of $100$ million inclusive $\psi^\prime$ MC events is used to
investigate possible backgrounds. No background events survive after
candidate selection.

Potential physics background contributions due to undetected or fake
photons and particle misidentification can come from the processes:
$\psi^\prime\rightarrow \pi^{0}p\bar{p}K^{+}K^{-}$,
$\psi^\prime\rightarrow\gamma\chi_{cJ}\rightarrow \gamma
K^{+}K^{-}K^{+}K^{-}$, $\gamma K^{+}K^{-}\pi^{+}\pi^{-}$, $\gamma
p\bar{p}\pi^{+}\pi^{-}$ and $\psi^\prime\rightarrow
p\bar{p}K^{+}K^{-}$. We produced $2\times10^{5}$ MC events for the
first process and $1\times10^{5}$ MC events for each of the other
processes in order to study these backgrounds. After applying the
event selection criteria to
MC events, $265$ events survive, and all of them are from
$\psi^\prime\rightarrow \pi^{0}p\bar{p}K^{+}K^{-}$. Since the
branching fraction of this channel has not been reported by the
PDG~\cite{PDG}, we determine it from
our data sample and use the result to estimate the background
contribution to be about $1.4$ events.  In addition, a $42.9$
pb$^{-1}$ data sample collected at $3.65$ \textrm{GeV} is used to
investigate possible continuum backgrounds, and no events survive
candidate selection.

\subsection{\boldmath $\chi_{cJ}\rightarrow p\bar{p}K^{+}K^{-}$}

\begin{figure*}[htbp]
   \centerline{
   \psfig{file=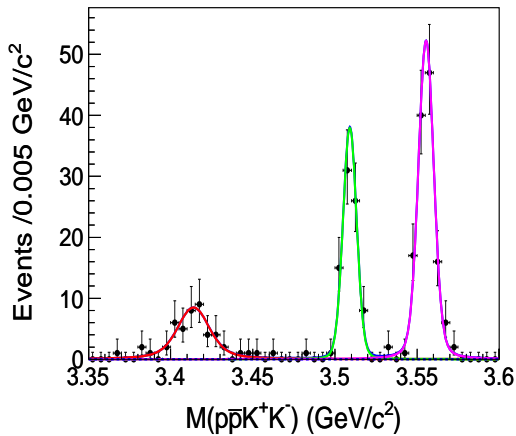,width=7cm,height=6cm, angle=0}
         }
         \caption{ The $p\bar{p}K^{+}K^{-}$ invariant mass
           distribution and fit result after excluding $\Lambda(1520),
           \bar{\Lambda}(1520)$ and $\phi$ intermediate states, where
           dots with error bars are data and the solid curves show the
           fit result. The dashed line, barely perceptible, is the
           estimated background component. }
   \label{ex:mass}
\end{figure*}

The branching fractions for $\chi_{cJ}\rightarrow p\bar{p}K^{+}K^{-}$
are measured excluding the evident $\Lambda(1520),\bar{\Lambda}(1520)$
and $\phi$ intermediate states seen in Fig.~\ref{fig:mass}(b,c,d), by vetoing these events with the mass
requirements $|M(pK^{-})-1.52|>0.07$ GeV/$c^2$,
$|M(\bar{p}K^{+})-1.52|>0.07$ GeV/$c^2$ and
$|M(K^{+}K^{-})-1.02|>0.03$ GeV/$c^2$.  The fit
of the invariant mass distribution of the remaining candidate events is
shown in Fig.~\ref{ex:mass}.

The $p\bar{p}K^+K^-$ mass distribution is fitted with Breit-Wigner
functions convolved with Gaussian resolution functions to describe the
$\chi_{cJ}$ signals and a flat distribution for the background, as
shown in Fig.~\ref{ex:mass}.  The Breit-Wigner parameters and Gaussian
instrumental resolutions are floated in the fit, with the $\chi_{cJ}$
widths fixed according to PDG~\cite{PDG} values.  The instrumental
resolutions are found to be about $4$ MeV/$c^2$, and the masses of the
$\chi_{cJ}$ from the fit are consistent with PDG values within
$1\sigma$.  The observed numbers of events, denoted as $N_{obs}$, for
$\chi_{cJ}\rightarrow p\bar{p}K^{+}K^{-}$ decays are listed in
Table~\ref{tab:ppkk}. The branching fractions are calculated according
to :

\begin{eqnarray*}
   \mathcal{B}(\chi_{cJ}\rightarrow p\bar{p}K^{+}K^{-})
      = \frac {N_{obs}} {N_{\psi^\prime} \cdot \mathcal{B}(\psi^\prime\rightarrow \gamma\chi_{cJ})
      \cdot \varepsilon },
\end{eqnarray*}
where $N_{\psi^\prime}$ is the total number of $\psi^\prime$ events,
which is measured to be $106\times10^{6}$ with an uncertainty of
$4\%$~\cite{Total}, the $\psi^\prime\rightarrow \gamma\chi_{cJ}$
branching fractions are taken from PDG~\cite{PDG} to be
$(9.62\pm0.31)\%$, $(9.2\pm0.4)\%$ and $(8.74\pm0.35)\%$ for
$\chi_{c0}$, $\chi_{c1}$ and $\chi_{c2}$, respectively, and the
detection efficiencies, $\varepsilon$, are determined
individually for simulated $\psi^\prime$ decays to
$p\bar{p}K^{+}K^{-}$ via the $\chi_{c0}$, $\chi_{c1}$ and $\chi_{c2}$
states.  The results are summarized in Table~\ref{tab:ppkk}.

\begin {table}[htp]
\begin {center}
\caption {The branching fractions for
          $\chi_{cJ}\rightarrow p\bar{p}K^{+}K^{-}$, where errors are statistical only.} \label{tab:ppkk}
\begin {tabular}{l>{\hspace{2pt}}c>{\hspace{2pt}}c>{\hspace{2pt}}c} \hline\hline
     Quantity   & $\chi_{c0}$  &  $\chi_{c1}$  &  $\chi_{c2}$    \\ \hline
     $N_{obs}$  & $48.2\pm7.7$ &  $81.5\pm9.2$ &  $131\pm12$     \\
     $\varepsilon(\%)$  & $3.8\pm0.1$ & $6.2\pm0.1$  & $6.8\pm0.1$  \\
     $\mathcal{B}(\chi_{cJ}\rightarrow p\bar{p}K^{+}K^{-})$ $(10^{-4})$ & $1.24\pm0.20$ &
                                   $1.35\pm0.15$ &
                                   $2.08\pm0.19$ \\\hline \hline
\end {tabular}
\end {center}
\end {table}

\subsection{\boldmath $\chi_{cJ}\rightarrow
     \bar{p}K^{+}\Lambda(1520)+\textbf{charge-conjugate~(c.c.)}$}

For analysis of the intermediate states described below, the following
invariant mass selection criteria are imposed:
\begin{center}
        $\chi_{c0}$ : $3.365$ GeV/$c^2$ $< M(p\bar{p}K^{+}K^{-}) <$ $3.455$ GeV/$c^2$

        $\chi_{c1}$ : $3.490$ GeV/$c^2$ $< M(p\bar{p}K^{+}K^{-}) <$ $3.530$ GeV/$c^2$

        $\chi_{c2}$ : $3.530$ GeV/$c^2$ $< M(p\bar{p}K^{+}K^{-}) <$ $3.580$ GeV/$c^2$
\end{center}

\begin{figure*}[htbp]
   \centerline{
   \psfig{file=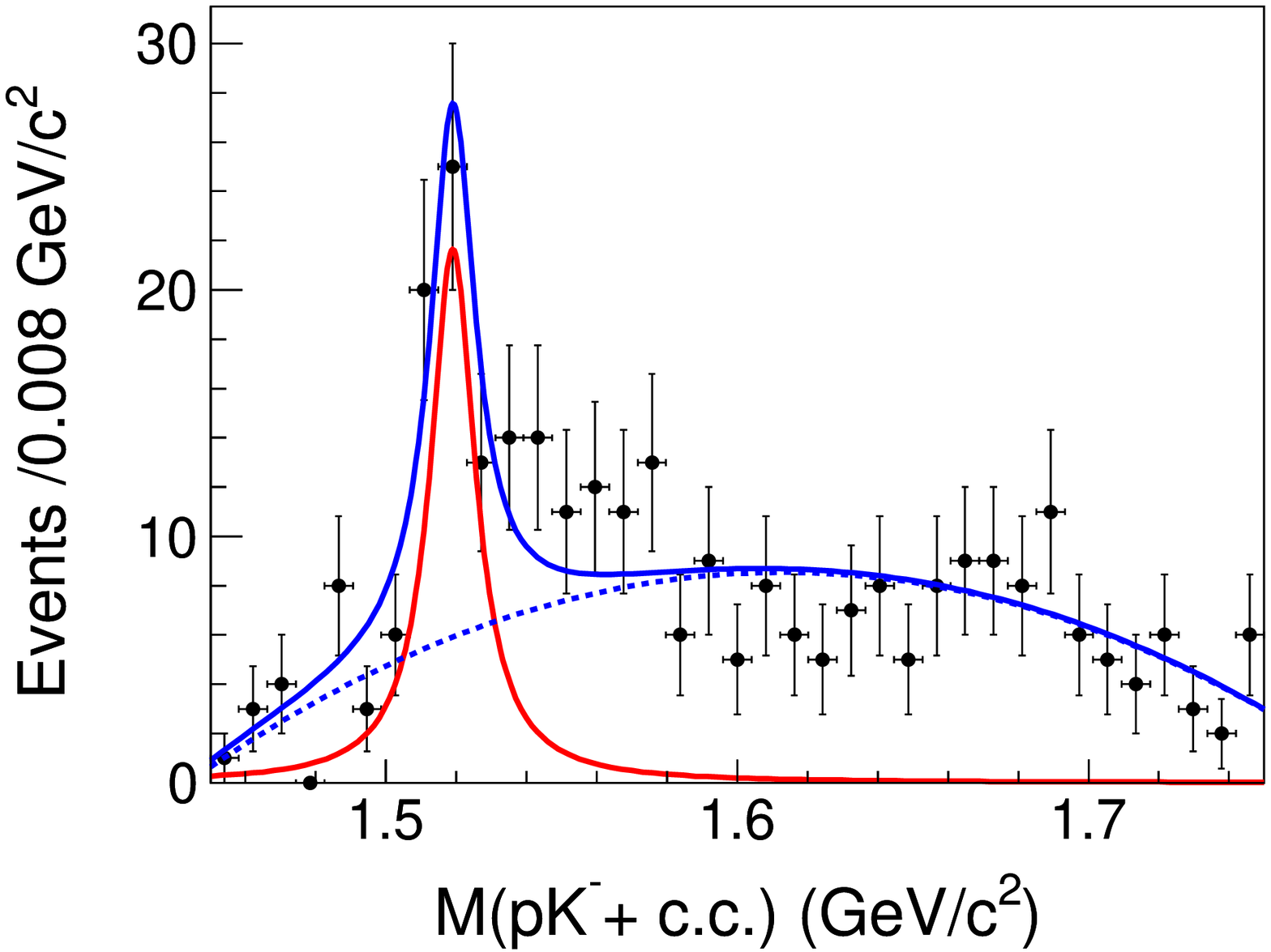,width=7cm,height=6cm, angle=0}
             \put(-35,135){(a)}
   \psfig{file=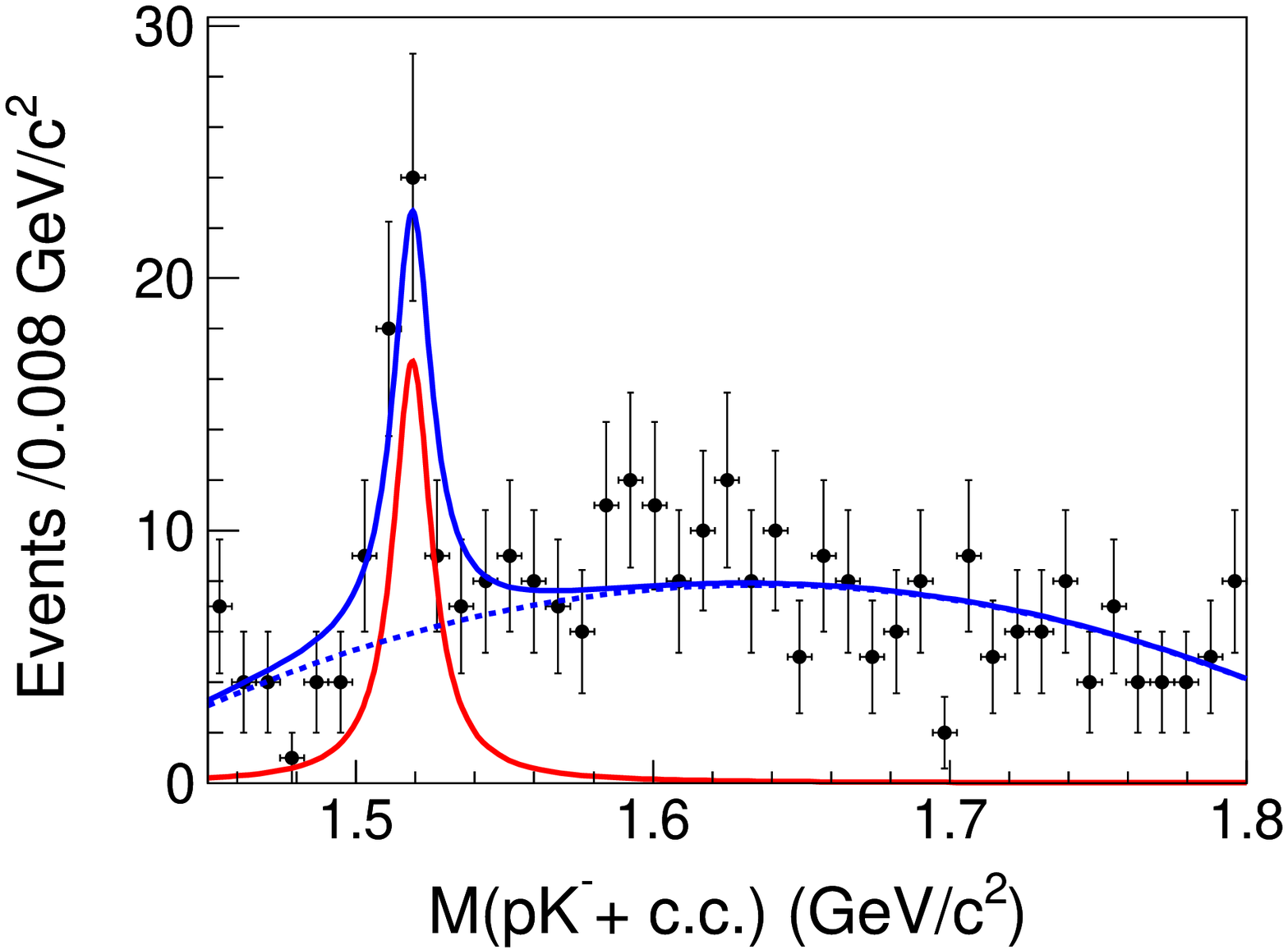,width=7cm,height=6cm, angle=0}
             \put(-35,135){(b)}
             }
      \centerline{
   \psfig{file=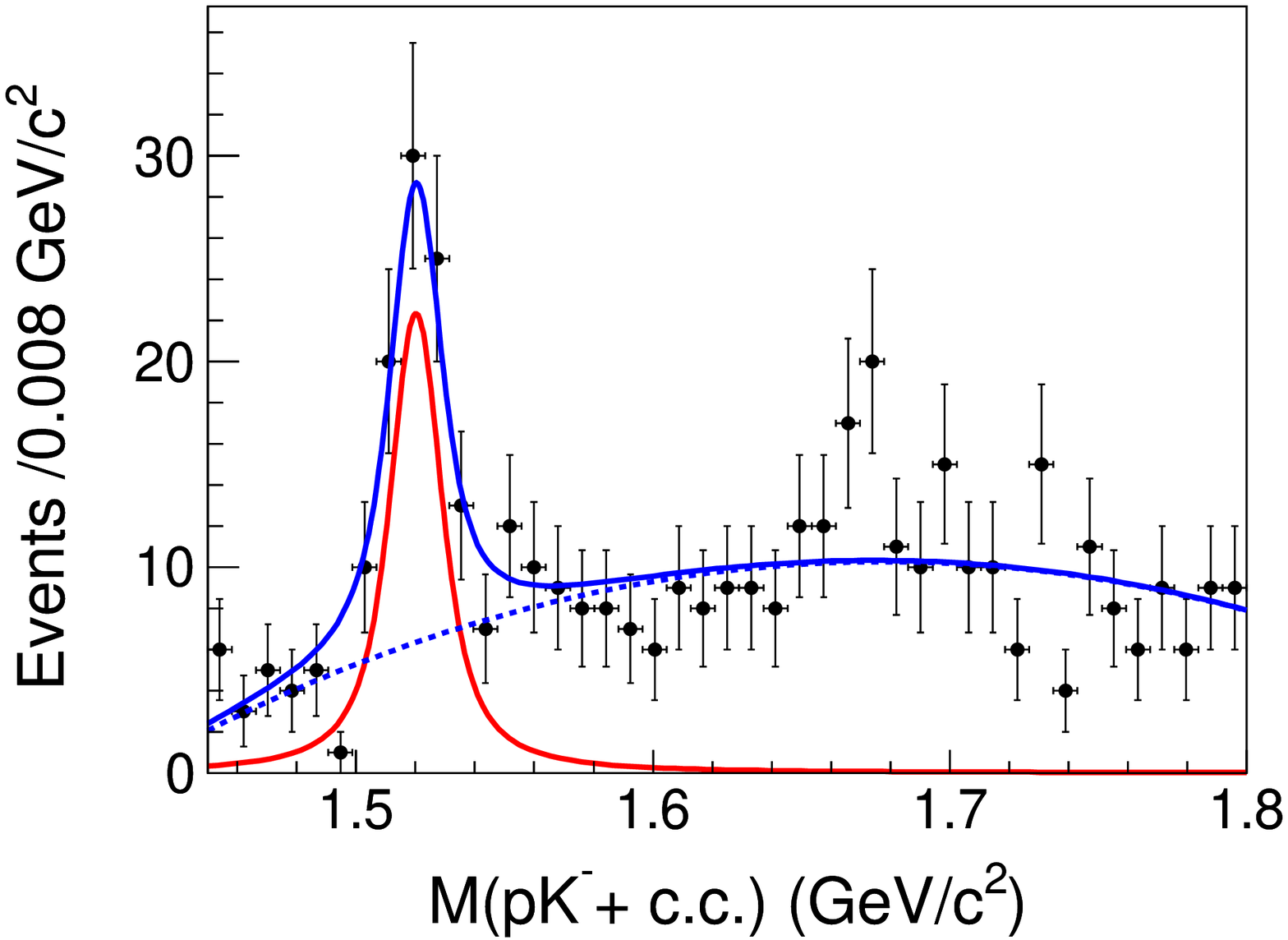,width=7cm,height=6cm, angle=0}
             \put(-35,135){(c)}
          }
   \caption{ Invariant mass distributions and fits to
             $pK^{-}+c.c$. in the decays of (a) $\chi_{c0}$, (b)
             $\chi_{c1}$ and (c) $\chi_{c2}$.  Dots with error bars
             are data. Solid lines are results of the fit, and dashed
             curves represent the background.  }
   \label{fig:lambda}
\end{figure*}

The three-body decay branching fractions $\chi_{cJ}\rightarrow
\bar{p}K^{+}\Lambda(1520)+c.c.$ are measured after rejecting the
$\bar{\Lambda}(1520)$ for $\bar{p}K^{+}\Lambda(1520)$ with the
requirement $|M(\bar{p}K^{+})-1.52|>0.07$ GeV/$c^2$ or for the
charge-conjugate $\Lambda(1520)$ to $pK^{-}\bar{\Lambda}(1520)$ by
$|M(pK^{-})-1.52|>0.07$ GeV/$c^2$. 
The fitted $pK^{-}+c.c$. invariant mass distributions are shown in
Fig.~\ref{fig:lambda}.  The fits use Breit-Wigner functions
convolved with Gaussians for the signals and Chebyshev polynomials
for backgrounds, where the Breit-Wigner masses and Gaussian
instrumental resolution are free parameters, and the widths of the
resonances are fixed to their PDG~\cite{PDG} values. The observed
numbers of events, $N_{obs}$, for $\chi_{cJ}\rightarrow
\bar{p}K^{+} \Lambda(1520)+c.c.$ are shown in Table~\ref{tab:lambda}.

The branching fractions are calculated according to:
\begin{eqnarray*}
   \mathcal{B}(\chi_{cJ}\rightarrow \bar{p}K^{+}\Lambda(1520)+c.c.)
      = \frac {N_{obs}} {N_{\psi^\prime} \cdot \mathcal{B}(\psi^\prime\rightarrow \gamma\chi_{cJ})
      \cdot \mathcal{B}(\Lambda(1520)\rightarrow pK^{-})  \cdot \varepsilon },
\end{eqnarray*}
where the detection efficiencies $\varepsilon$ are determined by MC
simulation of $\psi^\prime$ decays to $\bar{p}K^{+}\Lambda(1520)+c.c.$
for each of the $\chi_{c0}$, $\chi_{c1}$ and $\chi_{c2}$ states. The
results are summarized in Table~\ref{tab:lambda}.

\begin {table}[htp]
\begin {center}
\caption {The branching fractions for
          $\chi_{cJ}\rightarrow \bar{p}K^{+}\Lambda(1520)+c.c.$, where errors are statistical only.} \label{tab:lambda}
\begin {tabular}{l>{\hspace{2pt}}c>{\hspace{2pt}}c>{\hspace{2pt}}c} \hline\hline
     Quantity   & $\chi_{c0}$  &  $\chi_{c1}$  &  $\chi_{c2}$    \\ \hline
     $N_{obs}$  & $62\pm12$ &  $48\pm10$ &  $79\pm13$     \\
     $\varepsilon(\%)$  & $9.0\pm0.1$ & $12.1\pm0.1$  & $12.4\pm0.1$  \\
     $\mathcal{B}(\Lambda(1520)\rightarrow pK^{-})(\%)$ & $22.5$ & $22.5$  & $22.5$ \\
     $\mathcal{B}(\chi_{cJ}\rightarrow \bar{p}K^{+}\Lambda(1520)+c.c.)$ $(10^{-4})$ & $3.00\pm0.58$ &
                                   $1.81\pm0.38$ &
                                   $3.06\pm0.50$ \\\hline \hline
\end {tabular}
\end {center}
\end {table}

\subsection{\boldmath $\chi_{cJ}\rightarrow \Lambda(1520)\bar{\Lambda}(1520)$}

\begin{figure*}[htbp]
   \centerline{
   \psfig{file=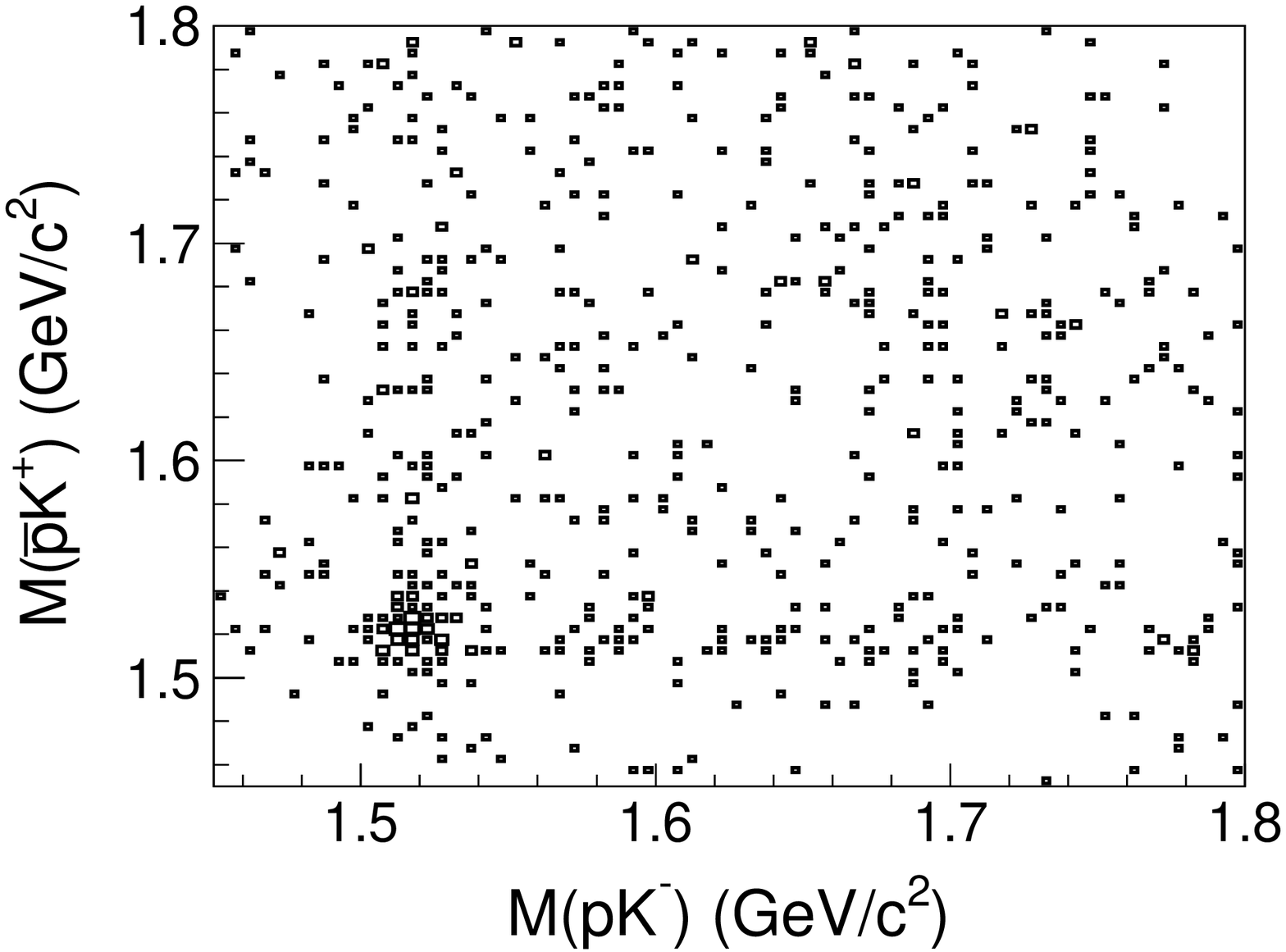,width=7cm,height=6cm, angle=0}
             \put(-35,135){(a)}
             \put(-98.8, 86){\color{red}\framebox(44,39)}
             \put(-94, 90){\large\bfseries C}
             \put(-98.8, 37){\color{red}\framebox(44,39)}
             \put(-94, 41){\large\bfseries B}
             \put(-156, 86){\color{red}\framebox(44,39)}
             \put(-151.2, 90){\large\bfseries A}
             \put(-156, 37){\color{blue}\framebox(44,39)}
             \put(-151.2, 41){\large\bfseries S}
   \psfig{file=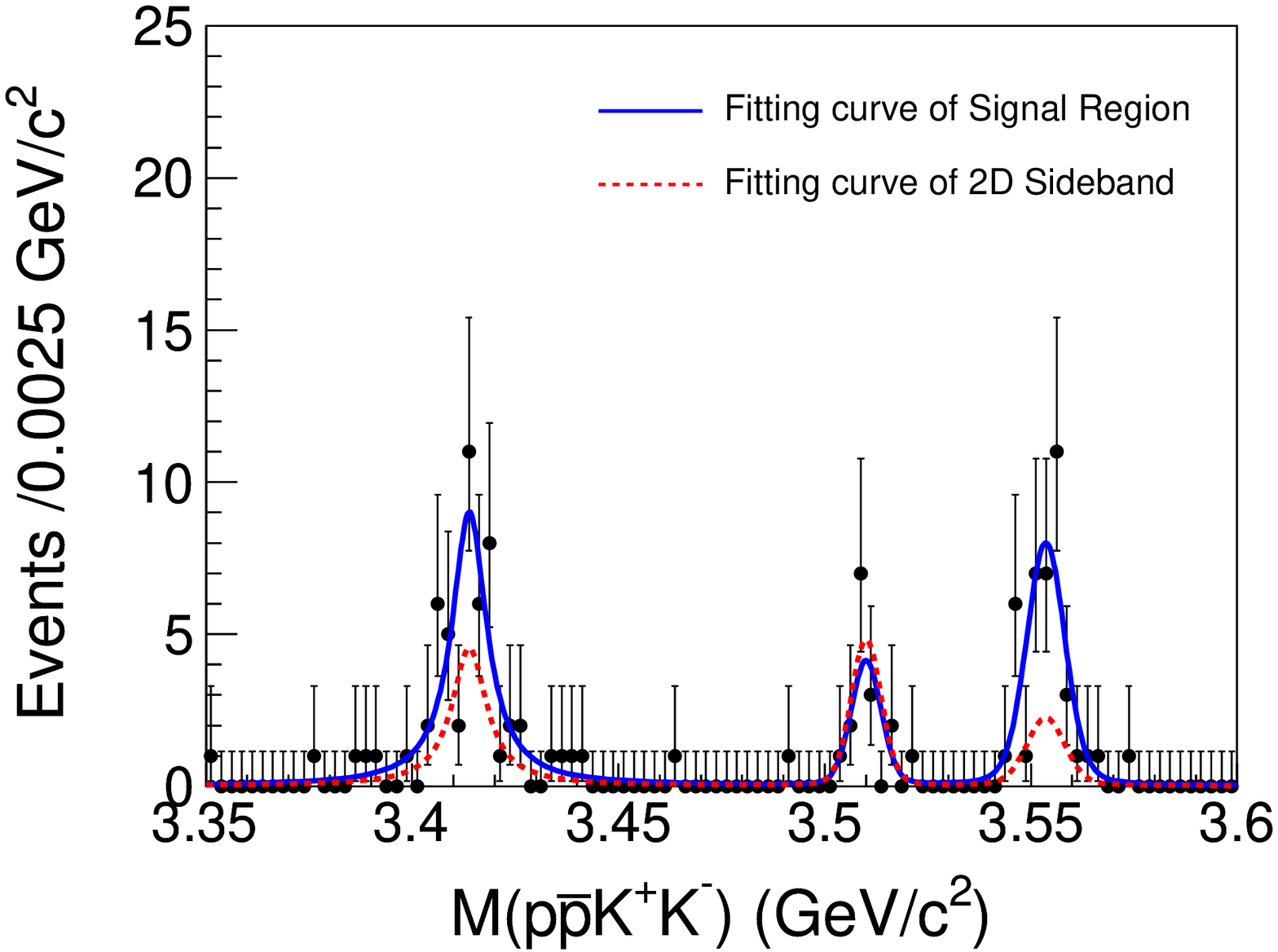,width=7cm,height=6cm, angle=0}
             \put(-145,135){(b)}
          }
   \caption{ (a) Scatter plot of $M(\bar{p}K^{+})$ versus $M(pK^{-})$;
     (b) Invariant mass spectrum and fits to $p\bar{p}K^{+}K^{-}$,
     where dots with error bars are events from the signal region. The solid line
     is the fitting curve for the events from signal region, and the
     dashed lines represent background estimated from the two-dimensional mass
     sidebands of regions ``A, B, C'' as shown in (a).}
   \label{fig:2lambda}
\end{figure*}

A scatter plot of the invariant mass $M(\bar{p}K^{+})$ versus
$M(pK^{-})$ is shown in Fig.~\ref{fig:2lambda}(a), where a signal for
$\chi_{cJ} \rightarrow \Lambda(1520) \bar{\Lambda}(1520)$ is evident.
Events remaining after rejecting $\phi p\bar{p}$ events with the veto
requirement $|M(K^{+}K^{-})-1.02|>0.03$ GeV/$c^2$ and
satisfying $|M(\bar{p}K^{+}) - 1.520| < 0.05$
GeV/$c^2$ and $|M(pK^{-}) - 1.520| < 0.05$ GeV/$c^2$
are selected as candidate $\Lambda(1520)\bar{\Lambda}(1520)$ events.
Two-dimensional
mass sideband regions used to estimate background in the signal region
S are indicated by the regions
A,B,and C in Fig.~\ref{fig:2lambda}(a). The events in these sideband regions
for data are scaled by factors that are determined by the
ratios of events in the signal region S to those in the
sideband regions for MC samples for the background channels, namely
$\psi^\prime\to \gamma\chi_{cJ}\to\gamma\bar{p}K^{+}\Lambda(1520)$,
$\psi^\prime\to\gamma\chi_{cJ}\to \gamma pK^{-}\bar{\Lambda}(1520)$
and $\psi^\prime\to\gamma\chi_{cJ}\to\gamma p\bar{p}K^{+}K^{-}$, are
to estimate the background in the signal region S of the data.

The mass spectra obtained from the signal and scaled sideband
background events in
Fig.~\ref{fig:2lambda}(a) are simultaneously fit using Breit-Wigner
functions convolved with Gaussian resolution functions. The
Breit-Wigner masses and the instrumental resolutions used for the
Gaussians are left as free parameters in the fit.  Other background is
described by a flat distribution.  The differences between the results
of the fits to the signal and scaled sideband events, shown in
Fig.~\ref{fig:2lambda}(b), are used to extract the
$\chi_{cJ}\rightarrow \Lambda(1520)\bar{\Lambda}(1520)$ yield.  We
find $28.1\pm9.8$ events for $\chi_{c0}\rightarrow \Lambda(1520)
\bar{\Lambda}(1520)$ and $28.9\pm7.4$ events for $\chi_{c2}\rightarrow
\Lambda(1520) \bar{\Lambda}(1520)$.  No distinct $\chi_{c1}\rightarrow
\Lambda(1520) \bar{\Lambda}(1520)$ signal is observed, and a $90\%$
C.L. upper limit is given using the Bayesian method.

The branching fractions are calculated according to:
\begin{eqnarray*}
\mathcal{B}(\chi_{cJ}\rightarrow \Lambda(1520) \bar{\Lambda}(1520))
~~~~~~~~~~~~~~~~~~~~~~~~~~~~~~~~~~~~~~~~~~~~~~~~~~~~~~~~~~~~~~~~~~~~~~~~~~~~~~~~~~ \\
~~~~~~~= \frac {N_{obs}} {N_{\psi^\prime} \cdot \mathcal{B}(\psi^\prime\rightarrow \gamma\chi_{cJ})
      \cdot \mathcal{B}(\Lambda(1520)\rightarrow pK^{-}) \cdot
      \mathcal{B}(\bar{\Lambda}(1520)\rightarrow \bar{p}K^{+})  \cdot \varepsilon }
\end{eqnarray*}
and the upper limit at the 90\% C.L. is calculated as:
\begin{eqnarray*}
\mathcal{B}(\chi_{c1}\rightarrow \Lambda(1520) \bar{\Lambda}(1520))
~~~~~~~~~~~~~~~~~~~~~~~~~~~~~~~~~~~~~~~~~~~~~~~~~~~~~~~~~~~~~~~~~~~~~~~~~~~~~~~~~~ \\
~~~~~~~< \frac {N_{obs}} {N_{\psi^\prime} \cdot \mathcal{B}(\psi^\prime\rightarrow \gamma\chi_{c1})
      \cdot \mathcal{B}(\Lambda(1520)\rightarrow pK^{-}) \cdot
      \mathcal{B}(\bar{\Lambda}(1520)\rightarrow \bar{p}K^{+})  \cdot \varepsilon \cdot (1-\sigma_{sys})},
\end{eqnarray*}
where the detection efficiencies are determined from MC simulation,
which assumes an angular distribution of $1+\alpha \cos^{2}\theta$ for
the two-body decays, and the value for $\alpha$ is estimated by
fitting the $\cos\theta$ distribution of data separately for the
$\chi_{c0}$, $\chi_{c1}$ and $\chi_{c2}$ states, $\theta$ is
the polar angle of a particle in the rest frame of its mother
particle, and $\sigma_{sys}$ denotes the systematic error (discussed
below). The results are summarized in Table~\ref{tab:2lambda}.

\begin {table}[htp]
\begin {center}
\caption {The branching fractions for
          $\chi_{cJ}\rightarrow \Lambda(1520) \bar{\Lambda}(1520)$. The errors are statistical only,
 and the upper limit is at the $90\%$ C.L.} \label{tab:2lambda}
\begin {tabular}{l>{\hspace{2pt}}c>{\hspace{2pt}}c>{\hspace{2pt}}c} \hline\hline
     Quantity   & $\chi_{c0}$  &  $\chi_{c1}$  &  $\chi_{c2}$    \\ \hline
     $N_{obs}$  & $28.1\pm9.8$ &  $<6.9$ &  $28.9\pm7.4$     \\
     $\varepsilon(\%)$  & $17.1\pm0.1$ & $16.3\pm0.1$  & $12.2\pm0.1$  \\
     $\mathcal{B}(\Lambda(1520)\rightarrow pK)(\%)$ & $22.5$ & $22.5$  & $22.5$ \\
     $\mathcal{B}(\chi_{cJ}\rightarrow \Lambda(1520)\bar{\Lambda}(1520))$ $(10^{-4})$ & $3.18\pm1.11$ &
                                   $<0.86$ &
                                   $5.05\pm1.29$ \\\hline \hline
\end {tabular}
\end {center}
\end {table}

\subsection{\boldmath $\chi_{cJ}\rightarrow p\bar{p}\phi$}

\begin{figure*}[htbp]
   \centerline{
   \psfig{file=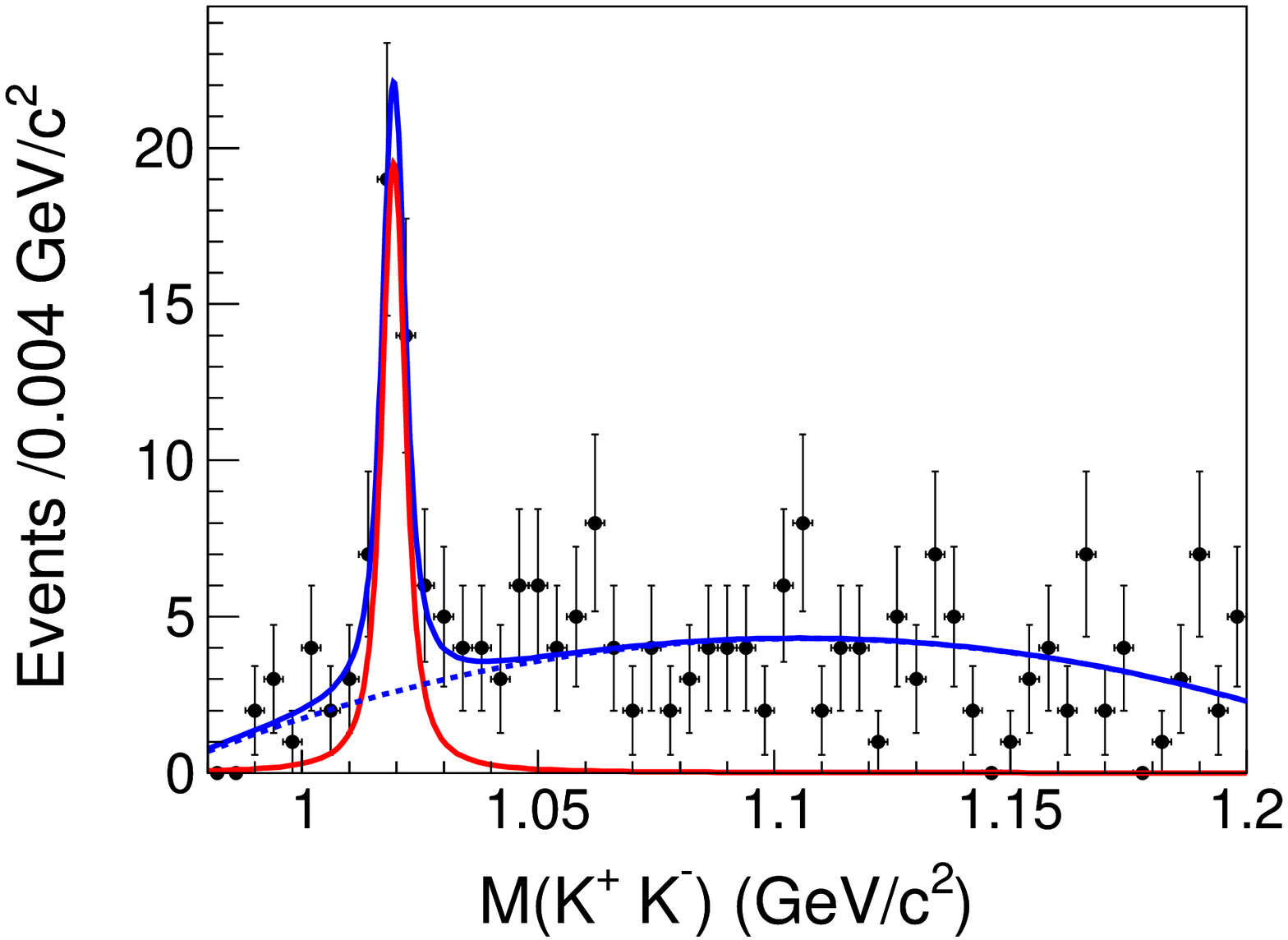,width=7cm,height=6cm, angle=0}
             \put(-35,135){(a)}
   \psfig{file=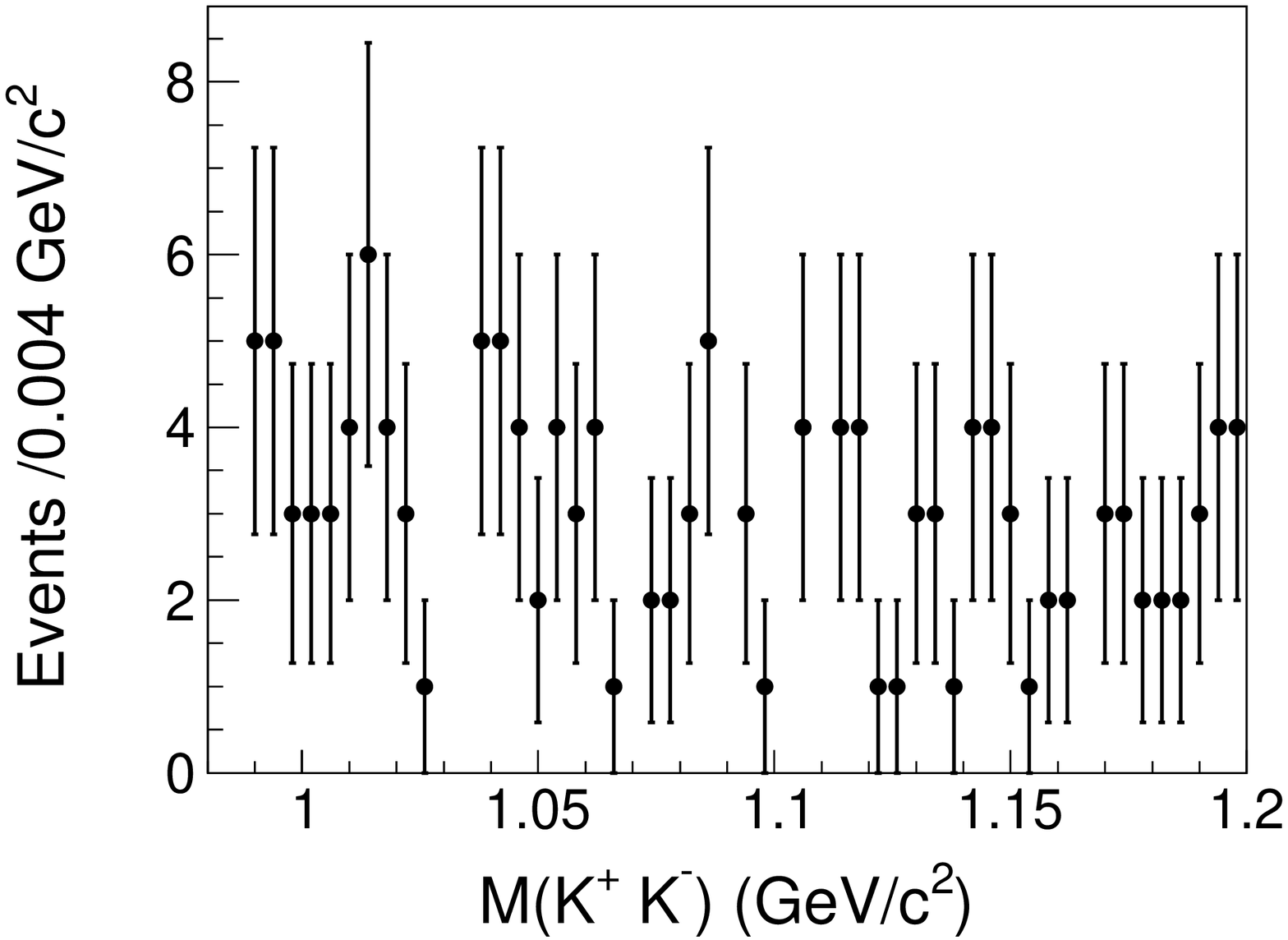,width=7cm,height=6cm, angle=0}
             \put(-35,135){(b)}
          }
   \psfig{file=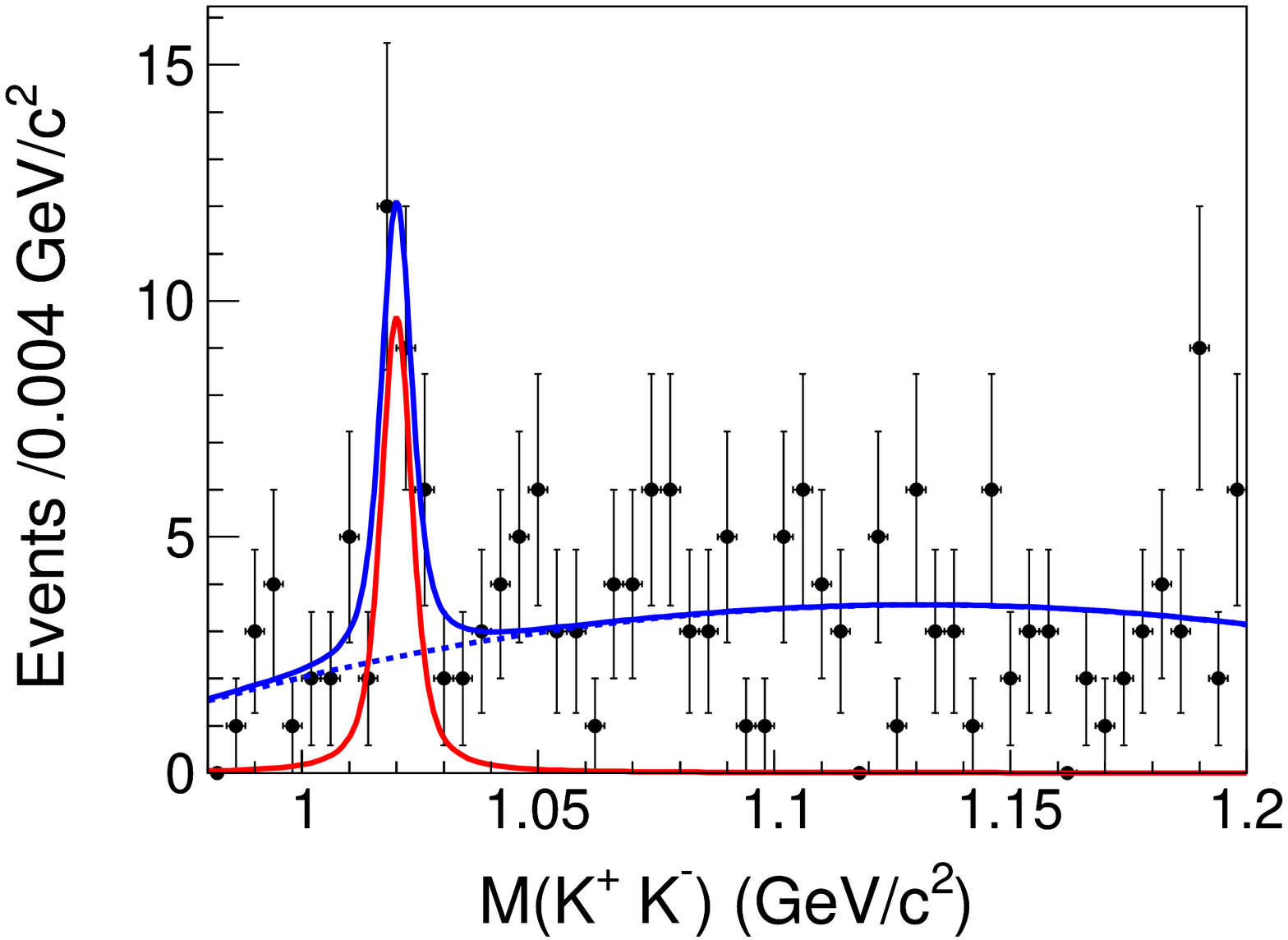,width=7cm,height=6cm, angle=0}
             \put(-35,135){(c)}
   \caption{$K^{+}K^{-}$ invariant mass distributions and fits to the
    spectra in the decays of (a) $\chi_{c0}$ and (c) $\chi_{c2}$. Dots
    with error bars are data, and solid lines represent the fit
    results. Dashed curves are background shapes. For decays of (b)
    $\chi_{c1}$, $\phi$ is not seen clearly, and the upper limit at
    the $90\%$ C.L. is given.}
   \label{fig:phi}
\end{figure*}

The $K^{+}K^{-}$ invariant mass distributions and fits to the spectra
are presented in Fig.~\ref{fig:phi} for the $\chi_{c0}$, $\chi_{c1}$,
and $\chi_{c2}$.  $\phi$ signals are observed clearly in the decays of
$\chi_{c0}$ (Fig.~\ref{fig:phi}(a)) and $\chi_{c2}$
(Fig.~\ref{fig:phi}(c)).  The fits use Breit-Wigner functions
convolved with Gaussians for the signals, where the Breit-Wigner
masses and instrumental resolutions are free parameters and resonance
widths are fixed at their PDG~\cite{PDG} values, and a Chebyshev
polynomial for the background. The observed numbers of events,
$N_{obs}$, for $\chi_{cJ}\rightarrow p\bar{p} \phi$ are listed in
Table~\ref{tab:phi}, as well as an upper limit at the 90\% C.L. for
$\chi_{c1}\rightarrow p\bar{p}\phi$ using a Bayesian method. The
branching fractions are estimated as:
\begin{eqnarray*}
   \mathcal{B}(\chi_{cJ}\rightarrow p\bar{p}\phi)
      = \frac {N_{obs}} {N_{\psi^\prime} \cdot \mathcal{B}(\psi^\prime\rightarrow \gamma\chi_{cJ})
      \cdot \mathcal{B}(\phi\rightarrow K^{+} K^{-})  \cdot \varepsilon }
\end{eqnarray*}
and the upper limit at the 90\% C.L. is calculated as:
\begin{eqnarray*}
\mathcal{B}(\chi_{c1}\rightarrow p\bar{p}\phi)
   < \frac {N_{obs}} {N_{\psi^\prime} \cdot \mathcal{B}(\psi^\prime\rightarrow \gamma\chi_{c1})
      \cdot \mathcal{B}(\phi\rightarrow K^{+} K^{-})  \cdot \varepsilon \cdot (1-\sigma_{sys})},
\end{eqnarray*}
where detection efficiencies are determined from MC simulation as
described above.
The results are summarized
in Table~\ref{tab:phi}.

\begin {table}[htp]
\begin {center}
\caption {The branching fractions for $\chi_{cJ}\rightarrow
          p\bar{p}\phi$. The errors are statistical only, and the
          upper limit is at the $90\%$ C.L..} \label{tab:phi}
\begin {tabular}{l>{\hspace{2pt}}c>{\hspace{2pt}}c>{\hspace{2pt}}c} \hline\hline
     Quantity   & $\chi_{c0}$  &  $\chi_{c1}$  &  $\chi_{c2}$    \\ \hline
     $N_{obs}$  & $42.4\pm8.2$ &  $<13.3$ &  $24.4\pm6.8$     \\
     $\varepsilon(\%)$  & $13.9\pm0.1$ & $17.7\pm0.1$  & $17.7\pm0.1$  \\
     $\mathcal{B}(\phi\rightarrow K^{+} K^{-})(\%)$ & $48.9$ & $48.9$  & $48.9$ \\
     $\mathcal{B}(\chi_{cJ}\rightarrow p\bar{p}\phi)$ $(10^{-5})$ & $6.12\pm1.18$ &
                                   $<1.58$ &
                                   $3.04\pm0.85$ \\\hline \hline\end {tabular}
\end {center}
\end {table}

\section{Systematic Uncertainty}

The main contributions to the branching fraction systematic
uncertainties originate primarily from the tracking, particle
identification, photon reconstruction, kinematic fit, branching
fractions of the intermediate states (from PDG~\cite{PDG}), total
number of $\psi^\prime$ events, fitting procedure and the event
generator.  The contributions of each item are summarized in
Table~\ref{sys:sum1} for
$\chi_{cJ}\rightarrow p\bar{p}K^{+}K^{-}$,
$\Lambda(1520)\bar{\Lambda}(1520)$ and Table~\ref{sys:sum2} for
$\bar{p}K^{+}\Lambda(1520)+c.c.$
and $p\bar{p}\phi$.

From analyses of very clean $J/\psi\rightarrow K^{\ast}K$ and
$J/\psi \rightarrow p\bar{p}\pi^{+}\pi^{-}$ decays, the
tracking efficiency for MC simulated events is found to agree with that
determined using data to within $2\%$ for each charged track. Hence,
$8\%$ is taken as the systematic uncertainty for the four charged
track final state.

The candidates of the selected final state require tracks be
identified as $p$, $\bar{p}$, $K^{+}$ or $K^{-}$. Comparing data and
MC event samples for $J/\psi \rightarrow \pi^{+} \pi^{-} p \bar{p}$
and $J/\psi\rightarrow K^{\ast}K$, a difference in MC and data
particle identification efficiency of 2\% is obtained for each
particle.  Hence, $8\%$ is taken as the systematic uncertainty for $p
\bar{p} K^{+} K^{-}$ identification.

Photon reconstruction efficiency is studied using $\psi^\prime\rightarrow
\pi^{+}\pi^{-} J/\psi \rightarrow\gamma \pi^{+} \pi^{-} p \bar{p}$, and the
difference between data and MC is about 1\% per photon~\cite{Total}.

To estimate the uncertainty from kinematic fitting, a
$\psi^\prime\rightarrow\gamma\chi_{cJ}\rightarrow\gamma p\bar{p}
\pi^{+}\pi^{-}$ sample is selected to study efficiency differences
between data and MC.  Errors of $1.4\%$, $1.6\%$ and $2.3\%$ are
obtained for decays of $\chi_{c0}$, $\chi_{c1}$ and $\chi_{c2}$,
respectively.

Uncertainties due to the decay model used in simulation for two-body and
three-body decay channels are estimated by varying the $\alpha$ values
in the decay angular distributions $1+\alpha \cos^{2}\theta$.  For
two-body decay channels, $\alpha$ is varied over a range such that the
angular distribution in MC is consistent with that of data. For
three-body decays, the accuracy of the angular distributions in data
are limited by low statistics. To be conservative, we vary $\alpha$
from $-1$ to $1$ and the resulting differences are taken as the systematic
uncertainty.

Uncertainties in the fitting procedure are obtained by altering
background shapes and fit intervals.  Uncertainties from the mass
window requirements, obtained by changing the requirements, of $\chi_{cJ}$,
$\Lambda(1520)$, $\bar{\Lambda}(1520)$ and $\phi$ are shown.

Uncertainties in the reconstruction efficiency for
$\chi_{cJ}\rightarrow p\bar{p}K^{+}K^{-}$ due to other possible
intermediate states, $\chi_{c1}\to\bar{p}K^{+}\Lambda(1600)+c.c.$ and
$\chi_{c0},~\chi_{c2}\to\bar{p}K^{+}\Lambda(1670)+c.c.$, which are not
pronounced in the data, are summarized in Table~\ref{sys:sum1}. Both
masses and widths of $\Lambda(1600)$ and $\Lambda(1670)$ are poorly
determined, and their branching fractions are not available. Their
branching fractions are taken conservatively as $5\times10^{-6}$, and
the systematic uncertainties are the differences between with and
without the intermediate states.

The total number of $\psi^\prime$ events with an uncertainty of 4\%
is obtained by studying inclusive hadronic $\psi^\prime$
decays~\cite{Total}.
The total systematic uncertainty is obtained by summing up
uncertainties contributed from all individual sources in quadrature.

\begin {table}[htp]
\begin {center}
\caption {Systematic uncertainties expressed in percent ($\%$) for the decay modes
          $\chi_{cJ}\rightarrow p\bar{p}K^{+}K^{-}$ and
          $\chi_{cJ}\rightarrow \Lambda(1520)\bar{\Lambda}(1520)$.
          } \label{sys:sum1}
\begin {tabular}{l|c c c c c c } \hline\hline
       &  \multicolumn{3}{p{0.3\textwidth}}{\centering $\chi_{cJ}\rightarrow p\bar{p}K^{+}K^{-}$}
       &  \multicolumn{3}{p{0.3\textwidth}}{\centering $\chi_{cJ}\rightarrow \Lambda(1520)\bar{\Lambda}(1520)$}  \\ \hline
     &   \multicolumn{1}{p{0.1\textwidth}}{\centering $\chi_{c0}$ } &
          \multicolumn{1}{p{0.1\textwidth}}{\centering $\chi_{c1}$ }    &
        \multicolumn{1}{p{0.1\textwidth}}{\centering $\chi_{c2}$ }  &
        \multicolumn{1}{p{0.1\textwidth}}{\centering $\chi_{c0}$ }   &
        \multicolumn{1}{p{0.1\textwidth}}{\centering $\chi_{c1}$ }   &
        \multicolumn{1}{p{0.1\textwidth}}{\centering $\chi_{c2}$ }  \\\hline
     Tracking  & $8.0$ & $8.0$ & $8.0$ & $8.0$ & $8.0$ & $8.0$      \\
     PID  &  $8.0$ & $8.0$ & $8.0$ & $8.0$ & $8.0$ & $8.0$ \\
     Photon recon. & $1.0$ & $1.0$ & $1.0$ & $1.0$ & $1.0$ & $1.0$ \\
     Kinematic Fit & $1.4$ & $1.6$ & $2.3$ & $1.4$ & $1.6$ & $2.3$ \\
     Fitting & $1.5$ & $0.6$ & $0.0$ & $0.0$ & $0.0$ & $0.0$ \\
     Mass window & $4.2$ & $6.0$ & $2.8$ & $8.2$ & --- & $11.0$ \\
     $\alpha$ value  &  --- & --- & --- & $3.3$ & --- & $4.1$ \\
     Branching fraction &  $3.2$ & $4.3$ & $4.0$ & $7.0$ & $7.6$ & $7.4$ \\
     $N_{\psi^\prime}$ &   $4.0$ & $4.0$ & $4.0$ & $4.0$ & $4.0$ & $4.0$ \\
     Efficiency & $5.0$ & $1.2$ & $5.9$ & --- & --- & ---   \\\hline
     Total & $14.2$ & $14.3$ & $14.5$ & $16.6$ & $14.3$ & $18.5$ \\\hline \hline
\end {tabular}
\end {center}
\end {table}

\begin {table}[htp]
\begin {center}
\caption {Systematic uncertainties expressed in percent ($\%$) for the decay modes
         $\chi_{cJ}\rightarrow \bar{p}K^{+}\Lambda(1520)+c.c.$ and
         $\chi_{cJ}\rightarrow p\bar{p}\phi$.
          } \label{sys:sum2}
\begin {tabular}{l|c c c c c c } \hline\hline
       &  \multicolumn{3}{p{0.3\textwidth}}{\centering $\chi_{cJ}\rightarrow \bar{p}K^{+}\Lambda(1520)+c.c.$}
       &  \multicolumn{3}{p{0.3\textwidth}}{\centering $\chi_{cJ}\rightarrow p\bar{p}\phi$}  \\ \hline
     &   \multicolumn{1}{p{0.1\textwidth}}{\centering $\chi_{c0}$ } &
          \multicolumn{1}{p{0.1\textwidth}}{\centering $\chi_{c1}$ }    &
        \multicolumn{1}{p{0.1\textwidth}}{\centering $\chi_{c2}$ }  &
          \multicolumn{1}{p{0.1\textwidth}}{\centering $\chi_{c0}$ }   &
        \multicolumn{1}{p{0.1\textwidth}}{\centering $\chi_{c1}$ }   &
        \multicolumn{1}{p{0.1\textwidth}}{\centering $\chi_{c2}$ }  \\\hline
     Tracking      & $8.0$ & $8.0$ & $8.0$ & $8.0$ & $8.0$ & $8.0$   \\
     PID           & $8.0$ & $8.0$ & $8.0$ & $8.0$ & $8.0$ & $8.0$   \\
     Photon recon. & $1.0$ & $1.0$ & $1.0$ & $1.0$ & $1.0$ & $1.0$   \\
     Kinematic Fit & $1.4$ & $1.6$ & $2.3$ & $1.4$ & $1.6$ & $2.3$   \\
     Fitting       & $9.4$ & $5.9$ & $6.8$ & $4.5$ &  ---  & $4.7$   \\
     Mass window   & $2.2$ & $3.6$ & $8.8$ & $2.1$ &  ---  & $1.0$   \\
     $\alpha$ value & $2.8$ & $2.6$ & $2.2$ & $4.0$ & $3.9$ & $2.5$   \\
     Branching fraction    & $5.4$ & $6.2$ & $5.9$ & $3.4$ & $4.4$ & $4.1$ \\
     $N_{\psi^\prime}$        & $4.0$ & $4.0$ & $4.0$ & $4.0$ & $4.0$ & $4.0$ \\\hline
     Total & $16.6$ & $15.5$ & $17.7$ & $14.1$ & $13.5$ & $14.0$ \\\hline \hline
\end {tabular}
\end {center}
\end {table}

\section{RESULTS AND DISCUSSION}

The measured branching fractions for the twelve decay modes decaying
to $p\bar{p}K^{+}K^{-}$ are summarized in
Table~\ref{tab:results}.  From the $106$ million $\psi^\prime$ decays
observed by BESIII at BEPCII, we report first measurements of these
branching fractions with uncertainties ranging from $20\%$ to $40\%$. With
larger statistics in future BESIII running, we expect to improve these
measurements and to be able to observe
$\Lambda(1520)\bar{\Lambda}(1520)$ in $\chi_{c1}$ decays.
The excited baryon $\Lambda(1520)\bar{\Lambda}(1520)$ decays provide new
information for evaluating model predictions of $\chi_{cJ}$ hadronic
decays.

\begin {table}[htp]
\begin {center}
\caption {Summary of branching fractions for twelve $\chi_{cJ}$ decay
  modes to $p\bar{p}K^{+}K^{-}$.  The first errors are
  statistical, and the second ones are systematic. The upper limits are at the $90\%$ C.L. including
  the systematic errors.} \label{tab:results}
\begin {tabular}{l>{\hspace{2pt}}c>{\hspace{2pt}}c>{\hspace{2pt}}c} \hline\hline
          & $\chi_{c0}$  &  $\chi_{c1}$  &  $\chi_{c2}$    \\ \hline
      $\mathcal{B}(\chi_{cJ}\rightarrow p\bar{p}K^{+}K^{-})$ $(10^{-4})$ & $1.24\pm0.20\pm0.18$ &
                                   $1.35\pm0.15\pm0.19$ &
                                   $2.08\pm0.19\pm0.30$ \\
      $\mathcal{B}(\chi_{cJ}\rightarrow \bar{p}K^{+}\Lambda(1520)+c.c.)$ $(10^{-4})$ & $3.00\pm0.58\pm0.50$ &
                                   $1.81\pm0.38\pm0.28$ &
                                   $3.06\pm0.50\pm0.54$ \\
      $\mathcal{B}(\chi_{cJ}\rightarrow \Lambda(1520)\bar{\Lambda}(1520))$ $(10^{-4})$ & $3.18\pm1.11\pm0.53$ &
                                   $<1.00$ &
                                   $5.05\pm1.29\pm0.93$ \\
     $\mathcal{B}(\chi_{cJ}\rightarrow p\bar{p}\phi)$ $(10^{-5})$ & $6.12\pm1.18\pm0.86$ &
                                   $<1.82$ &
                                   $3.04\pm0.85\pm0.43$ \\\hline \hline
\end {tabular}
\end {center}
\end {table}

\section{Acknowledgments}

The BESIII collaboration thanks the staff of BEPCII and the
computing center for their hard efforts. This work is supported in
part by the Ministry of Science and Technology of China under
Contract No. 2009CB825200; National Natural Science Foundation of
China (NSFC) under Contracts Nos. 10625524, 10821063, 10825524,
10835001, 10875113, 10935007, 10979038, 11005109, 11079030;
the Chinese Academy of Sciences (CAS)
Large-Scale Scientific Facility Program; CAS under Contracts Nos.
KJCX2-YW-N29, KJCX2-YW-N45; 100 Talents Program of CAS;
Research Fund for the Doctoral Program of Higher Education of China
under Contract No. 20093402120022; Istituto
Nazionale di Fisica Nucleare, Italy;
Siberian Branch of Russian Academy of Science, joint project No 32
with CAS; U. S. Department of Energy under Contracts Nos.
DE-FG02-04ER41291, DE-FG02-91ER40682, DE-FG02-94ER40823; University
of Groningen (RuG) and the Helmholtzzentrum fuer
Schwerionenforschung GmbH (GSI), Darmstadt; WCU Program of National
Research Foundation of Korea under Contract No.
R32-2008-000-10155-0.



\end{document}